\documentclass[a4paper,11pt]{article}
\usepackage{pos}
\usepackage{ptdr-definitions}
\usepackage{hepparticles}
\usepackage{heppennames2}

\usepackage{amsmath}
\begin{document}
\title{CMS Results at the t\textoverline{t} threshold}
\author*[a]{J\"orn Bach}
\onbehalf{on behalf of the CMS Collaboration}
\affiliation[a]{Deutsches Elektronensynchrotron DESY,\\ Notkestr. 85, 22761 Hamburg, Germany }
\emailAdd{joern.bach@desy.de}
\abstract{Recent result of CMS at the top-antitop quark pair production threshold in the dileptonic and semileptonic decay channels are presented. The results were obtained analyzing 138 fb$^{-1}$ of pp collision data taken at $\sqrt{s} = 13$ TeV with the CMS detector. Spin sensitive variables are combined with the t\textoverline{t} invariant mass $m_\text{t\textoverline{t} }$ to achieve a high sensitivity for intermediate pseudoscalar and scalar states. An excess in the data is observed for low values of $m_\text{t\textoverline{t}}$, favoring a pseudoscalar over a scalar hypothesis. The results are given in the interpretation of a simplified t\textoverline{t} bound state model {$\eta_t$} and in a generic model for heavy (pseudo)scalar production. The observed excess is compatible with the non-relativistic QCD (NRQCD) calculations for a t\textoverline{t} bound state.}
\FullConference{The Thirteenth Annual Large Hadron Collider Physics (LHCP2025)\\
5-9 May 2025\\
Taipei
}
\maketitle
\section{Introduction} 
Top-Antitop quark pair production and its subsequent decays offer a promising opportunity to research beyond the standard model (BSM) theories. 
In particular, models with particles featuring  Yukawa-like couplings exhibit a big discovery potential here as the coupling is largest due to the high top quark mass.
Many models that fit these criteria include extended Higgs sectors, such as 2 Higgs Doublet Models (2HDM), which also occur in the context of Supersymmetry. A key feature of these models is - among other new particles - the presence of heavy pseudoscalar (A) and scalar Higgs boson (H). For large parts of the extended Higgs parameter space these two particles dominantly decay to a top quark-antiquark pair.  
In addition to BSM signatures, for \ttbar masses close to 2*$m_\text{t}$, there is a potential contribution from bound states, predicted in earlier calculations \cite{Kiyo:2008bv, Sumino:2010bv, Ju:2020otc, Fadin:1987wz, Fadin:1990wx}.
The top quark bound state effects are expected to form a broad peak below the t\textoverline{t} threshold and exhibit dominantly pseudoscalar behavior at the LHC. They have not previously been observed. 
With these two signatures being very similar, a search for both BSM and SM scenarios is based on the same methodologies. 
This work presents such a search \cite{CMS-HIG-22-013}, using the full Run 2 dataset of the CMS detector ~\cite{Chatrchyan:2008zzk, CMS:2023gfb}  with an integrated luminosity of 138 fb$^{-1}$. This work superseedes the previous CMS search for A and H bosons with only 35.9 fb$^{-1}$ of data \cite{CMS:2019pzc}. The results consider both the semileptonic (\ttlj) decay channel and the dileptonic (\ttll) decay channel. For the threshold \ttbar bound state results in \cite{CMS:2025kzt}, exclusively the dileptonic decays are considered. 

\section{Signal Modeling}
The heavy Higgs A and H scenarios are considered to be produced by gluon fusion in a top quark loop. The Yukawa-like couplings are implemented as coupling modifiers (g$_\text{At\textoverline{t}}$ and g$_\text{Ht\textoverline{t}}$). Additionally, the mass and width are set to be free parameters. 
Since the final state is indistinguishable from the SM t\textoverline{t}, an interference term arises. This leads to a characteristic peak-dip structure in the $m_\text{t\textoverline{t}}$ spectrum.\\ 
For t\textoverline{t} bound states there is no full Monte-Carlo calculation available that can be compared to data. In literature \cite{Kiyo:2008bv} there are NRQCD calculations available for the expected $m_\text{t\textoverline{t}}$ spectrum. 
The results have two important contributions, one from a color singlet that forms an attractive potential and thus leads to a peak below the t\textoverline{t} threshold. The other contribution comes from the color octet component. It is repulsive and has only a small impact at the threshold. To compare these predictions to data, a simplified model $\eta_\text{t}$ \cite{Fuks:2021xje, Maltoni:2024tul} is constructed that models the peak below the threshold. The $\eta_\text{t}$ is a generic spin-0, color-singlet pseudoscalar resonance with direct couplings to gluons and top quarks. The mass and width are derived from a fit to the NRQCD calculation. The mass is $m(\eta_\text{t})$ = 343 GeV and the width is $\Gamma(\eta_\text{t})$ = 7 GeV. 
This model is added to the perturbative QCD continuum (pQCD) prediction. The $\eta_\text{t}$ is not a full representation of the actual lineshape of t\textoverline{t} bound state effects but this is considered sufficient for this analysis as the $m_\text{t\textoverline{t}}$ resolution is 15\% in the threshold region.

\section{Analysis Setup}
In this talk we focus on the \ttll channel since it is most relevant for the results at the threshold.
The selection for the dileptonic channel requires exactly two opposite sign leptons (e/$\mu$) and at least two jets, with one or more of them b-tagged. Three categories in ee, $\mu\mu$ and e$\mu$ are constructed to apply an additional $p_\text{T}^\text{miss} > $ 40 GeV cut in the same-flavor categories. Additionally, the Z-peak is cut away everywhere to reduce the influence of Z+jets events. 
Then, the t\textoverline{t} four-momenta are reconstructed with an analytical algorithm\cite{Sonnenschein:2006ud}. The inputs to this calculation are smeared to the detector resolution and the reconstruction is repeated 100 times taking the weighted average as a result. The algorithm assumes all missing transverse momentum $p_\text{T}^\text{miss}$ to be from the neutrinos and the top quarks and W boson to be on-shell.
For the statistical analysis, three-dimensional templates are constructed from 20 bins in the m$_\text{t\textoverline{t}}$ observable as well as 3 bins each in the two spin-correlation observables c$_\text{hel}$ and c$_\text{han}$. They are defined as the scalar product of the unit lepton momenta in their respective parent top frames. For c$_\text{han}$ the component of one lepton momentum is flipped in direction of the top quark direction of flight. These observables probe the t\textoverline{t} spin density matrix components sensitive to the alignment between the top and antitop quark spins \cite{Bernreuther:2015yna}. For pure $^1\text{S}_0$ states, such as produced by pseudoscalars A and $\eta_\text{t}$, the slope is maximally positive in c$_\text{hel}$. For pure $^3\text{P}_0$ states, such as produced by the scalar H, the slope of c$_\text{han}$ is maximally negative. By employing both, a strong discrimination between the different signal hypothesis can be reached. \\ 
The background arises dominantly from SM t\textoverline{t} production which is modeled with \POWHEGhvqPYTHIA in NLO and corrected to NNLO accuracy by a 2D reweighting in m$_\text{t\textoverline{t}}$ and cos$\theta^*$. Further backgrounds are single top production and W associated single top production. Both are computed in MC. Z+jets production is also modeled in MC and normalized with a data driven method. 

\section{Results}
\begin{figure}
	\includegraphics[width=\textwidth]{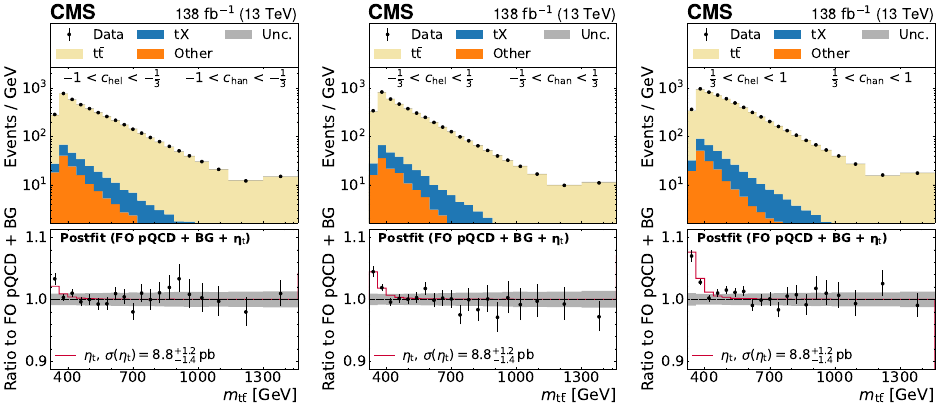}
	\caption{Observed data (black points) and fitted Monte Carlo prediction. An excess can be observed  near the t\textoverline{t} threshold. The observed distribution best fits the $\eta_\text{t}$ hypothesis for a cross section of $\sigma_{\eta_\text{t}}$ = 8.8 $^{+1.2}_{-1.4}$ pb. The figure is taken from \cite{CMS:2025kzt}.}
	\label{fig:res1}
\end{figure}
The \mtt distribution shows a strong excess near the top-antitop quark pair production threshold. The excess is most pronounced in the highest c$_\text{hel}$ and c$_\text{han}$ bin. Figure \ref{fig:res1} shows three bins in c$_\text{hel}$ and c$_\text{han}$ from low values to high values. 
The observed data is consistent with the $\eta_\text{t}$ prediction for a best fit value of the cross section $\sigma_{\eta_\text{t}}$ = 8.8 $^{+1.2}_{-1.4}$ pb. This exceeds five standard deviations from the zero hypothesis. Comparing the extracted cross section to the NRQCD theory prediction of $\sigma_{\eta_\text{t}}^\text{pred}$ = 6.43 pb \cite{Fuks:2021xje}, showing good agreement.
The search is dominated by systematic uncertainties. The leading systematic uncertainties are related to \ttbar modeling. This includes generator uncertainties such as the difference of \POWHEG \hvq to \POWHEG \bbfourl and \PYTHIA to \HERWIG which are implemented as shape uncertainties of the difference of the templates. Also the parton shower final state radiation $\alpha_S$ and the top mass uncertainties have high impacts on the result.  

\begin{table}[!ht]
\centering
\renewcommand{\arraystretch}{0.9}
\begin{tabular}{cc}
    \hline
    FO pQCD generator setup & \etatxsec [pb]{\Large\strut} \\
    \hline
    \POWHEGhvqPYTHIA & $8.7 \pm 1.1$ \\
    \POWHEGhvqHERWIG & $8.6 \pm 1.1$ \\
    \MGvATNLOFXFXPYTHIA & $9.8 \pm 1.3$ \\
    \POWHEGbbfourlPYTHIA & $6.6 \pm 1.4$ \\[\cmsTabSkip]
    Nominal result & \etatxsecval{\Large\strut} \\
    \hline
\end{tabular}
\caption{%
    Results for \etatxsec extracted with different simulated FO pQCD {\ttbar}(+\tW) predictions.
    Nuisance parameters encoding generator difference uncertainties are not included in these likelihood fits.
    The nominal result, \ie, \POWHEGhvqPYTHIA, includes these nuisance parameters for comparison. The table is taken from \cite{CMS:2025kzt}.
}
\label{tab:generators}
\end{table}

Since the search is very sensitive to the modeling of the t\textoverline{t} MC prediction, further sanity checks with other MC generators are carried out. For this, the nominal t\textoverline{t} model (including shape uncertainties of \POWHEG \bbfourl and \HERWIG) is exchanged for \POWHEG\hvq interfaced with \PYTHIA, \POWHEG \hvq interfaced with \HERWIG, \MGvATNLO FxFx interfaced with \PYTHIA and \POWHEG \bbfourl interfaced with \PYTHIA. These alternative fits do not include shape uncertainties for t\textoverline{t} MC generators. The cross section results are presented in table \ref{tab:generators}. They are compatible with each other within 1.5 standard deviations. The lower extracted cross section of the \POWHEG \bbfourl setup is caused by a slope in the c$_\text{hel}$ observable at the t\textoverline{t} threshold, resulting in a smaller excess with respect to the nominal result. The excess remains significant across all scenarios.   
To establish the pseudoscalar nature of the excess, a two dimensional fit was done that included a scalar $\chi_\text{t}$ which was constructed similarly to the $\eta_\text{t}$. The extracted $\chi_\text{t}$ cross section of $\sigma(\chi_\text{t})$ = 3.0$^\text{+2.6}_\text{-3.3}$pb is compatible with 0 within one standard deviation while the $\sigma_{\eta_\text{t}}$ = 7.8 $^{+1.8}_{-1.2}$ pb still exceeds five standard deviations from zero.\\
\begin{figure}
	\begin{minipage}{0.49\textwidth}
		\includegraphics[width=\textwidth]{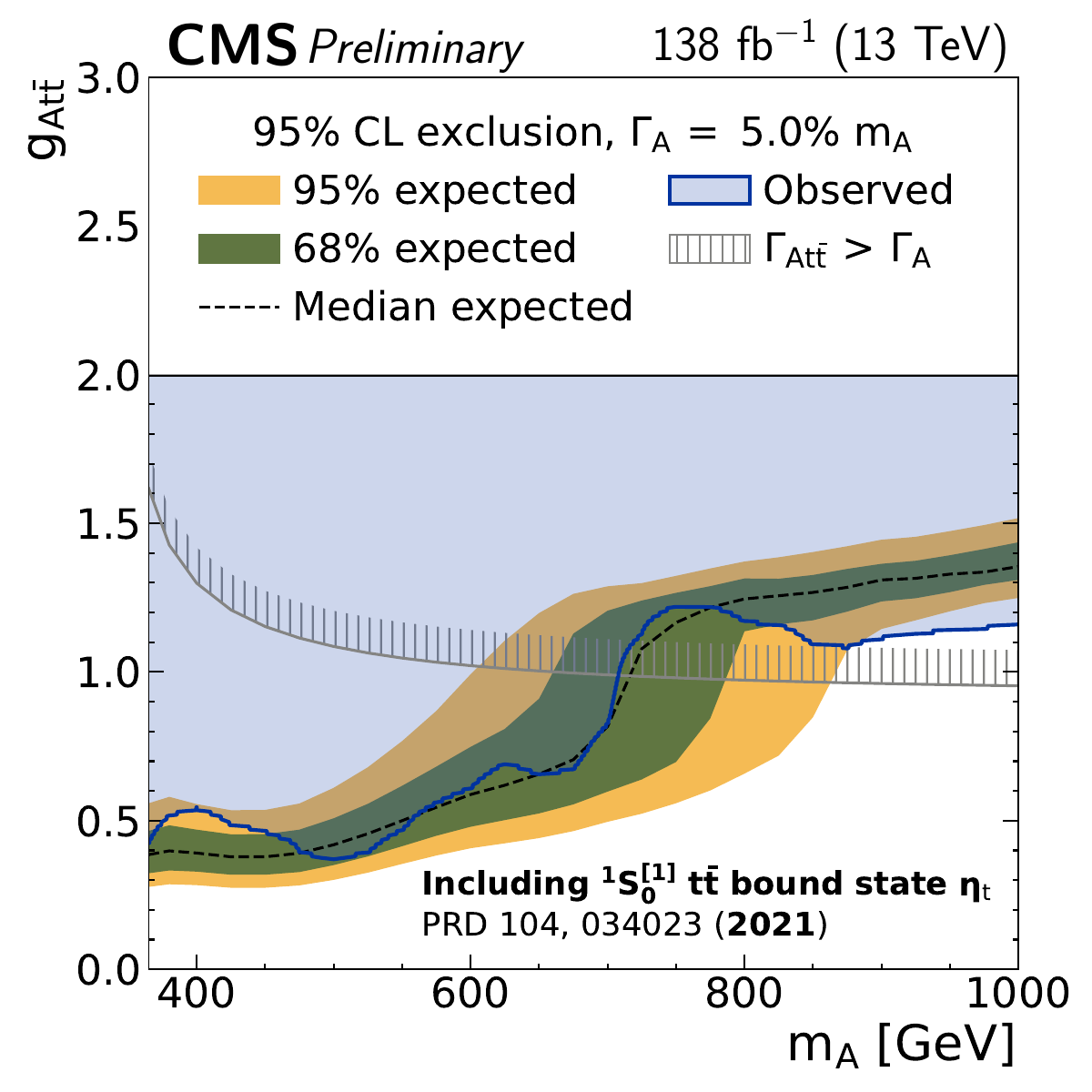}
	\end{minipage}
	\begin{minipage}{0.49\textwidth}
		\includegraphics[width=\textwidth]{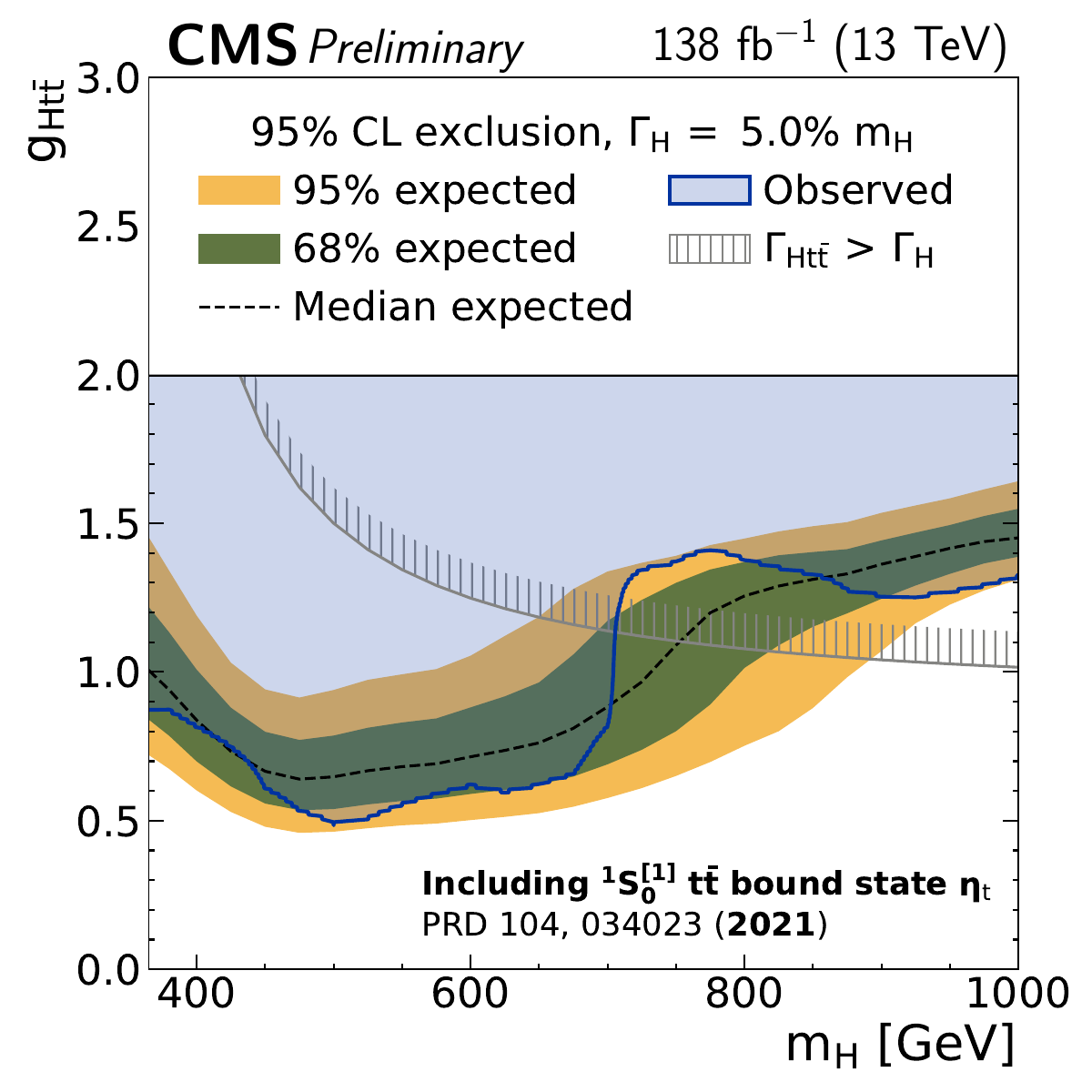}
	\end{minipage}
	\caption{Exclusion limits for the pseudoscalar A (left) and scalar H (right) scenario with relative width of $\Gamma_\text{A/H}$ = 5\% with the \etat model in the background with a floating normalization. These limits include both the \ttll and the \ttlj channels. Figure taken from \cite{CMS-HIG-22-013}.}
	\label{fig:ah}
\end{figure}
The \etat interpretation describes the excess well. Subsequently, limits on the A and H hypothesis are set with \etat in the background. For this, both the \ttll and the \ttlj channel are used. 
The normalization of the \etat contribution is a free parameter of the fit.
The \etat absorbs the excess fully and no significant deviations from the SM prediction are found. Figure \ref{fig:ah} shows two exemplary parameter scans for A and H hypotheses with a width of $\Gamma_\text{A/H}$ = 5\%.  Additionally to the one dimensional results, two dimensional scans in \gAtt and \gHtt simultaneously are also given.These contours are computed for several mass and width hypothesis and can be interpreted in a 2HDM model of choice because of their generic nature.\\ 
It has to be noted that the 15\% resolution in \mtt at the \ttbar threshold does not allow us to discriminate A and $\eta_t$ hypotheses, leaving room for a BSM contribution in this sector
In fact, more complex descriptions of \ttbar bound state effects beyond the simplified \etat model can not be resolved at this point as well. 

\acknowledgments J.B. acknowledges DESY, a member of the Helmholtz Association, and DASHH, Data Science in Hamburg - Helmholtz Graduate School  for the Structure of Matter, for financial support.

\bibliographystyle{JHEP}
\bibliography{thresholdbib}

\end{document}